\documentstyle[amssymb,prabib,preprint,aps,psfig,graphicx]{revtex}

\begin{document}
\title{Vacuum polarization in the spacetime of charged \\
nonlinear black hole}
\author{Waldemar Berej\thanks{
Electronic Address: berej@tytan.umcs.lublin.pl} and Jerzy Matyjasek\thanks{
Electronic Address: matyjase@tytan.umcs.lublin.pl, jurek@kft.umcs.lublin.pl}}
\address{Institute of Physics, Maria Curie-Sk\l odowska University,\\
pl. Marii Curie-Sk\l odowskiej 1,\\
20-031 Lublin, Poland}
\maketitle
\tighten

\begin{abstract}
Building on general formulas obtained from the approximate
renormalized effective action, the approximate stress-energy
tensor of the quantized massive scalar field with arbitrary
curvature coupling in the spacetime of charged black hole
being a solution of coupled equations of nonlinear
electrodynamics and general relativity is constructed and
analysed. It is shown that in a few limiting cases, the
analytical expressions relating obtained tensor to the
general renormalized stress-energy tensor evaluated in the
geometry of the Reissner-Nordstr\"{o}m black hole could be
derived. A detailed numerical analysis with special emphasis
put on the minimal coupling is presented and the results are
compared with those obtained earlier for the conformally
coupled field.  Some novel features of the renormalized
stress-energy tensor  are discussed.

\end{abstract}


\preprint{}

\section{Introduction}

\label{intro}

One of the most intriguing open questions in modern theoretical physics is the issue of
a final stage of the black hole evaporation. Although a definite answer would
be possible only with the full machinery of quantum theory of gravity, some
important preliminary results may be, in principle, obtained within the
framework of semiclassical theory. However, as the semiclassical approach
could not be used to describe the evolution of the system in the Planck regime,
it can, at best, tell us about a tendency rather than the limit itself.
Unfortunately, even this simplified programme is hard to execute as the
semiclassical Einstein field equations comprise a rather complicated set of
nonlinear partial differential equations, and, moreover, the source term ---
the renormalized stress-energy tensor --- should be known for a wide class
of nonstatic metrics. It is natural, therefore, that in order to make the back
reaction problem tractable,  one should refer to some approximations. 

It seems that the most promising approach consists in constructing the
approximate stress-energy tensor in the geometry of a static black hole and
subsequent computation of the semiclassical corrections to the classical
metric. Although evaluated in the static background, such corrections, as
has been pointed out in Ref.~\cite{THA00}, are relevant because
they give direct
information about the influence of the quantum effects on the temperature
of a black hole. Thus far this programme, initiated in Ref. \cite{York:1985wp},
has been carried out for massless fields in Schwarzschild spacetime 
\cite{Lousto:1988sp,ho1,Hochberg:1993xt,Hochberg:1994ps,Anderson:1994hh,Kocio} and for
massive scalar fields with arbitrary curvature coupling in
Reissner-Nordstr\"{o}m (RN) geometry~\cite{THA00}, where, among other things,
quantum corrections to the geometry, entropy, and trace anomaly were computed.
The most important ingredient of the approach is therefore the renormalized
stress-energy tensor of the quantized field propagating in a spacetime of
the static black hole constructed in a physically interesting state. However, direct evaluation
of this object  is rather complicated.

It is
believed that the physical content of the theory of quantum fields in curved
spacetime is contained in the (renormalized) effective action, $W_{R},$ a
useful quantity allowing evaluation of the stress-energy tensor by means of
the standard formula 
\begin{equation}
{\frac{2}{g^{1/2}}}{\frac{\delta }{\delta g_{ab}}}W_{ren}\,=\,\langle
T^{ab}\rangle _{ren}.  \label{stress}
\end{equation}
Unfortunately, the effective action is a nonlocal functional of the metric
and its exact form is unknown. In the attempts to construct the renormalized
stress-energy tensor one is forced, therefore, to employ numerical methods or
to accept some approximations.

For the quantized massive fields in the large mass limit, i.e. when the
Compton length is much smaller than the characteristic radius of a
curvature, the nonlocal contribution to the effective action can be
neglected, and the series expansion in $m^{-2}$ of the renormalized
effective action, $W_{R},$ may be constructed with the aid of the
DeWitt-Schwinger method~\cite{MR83j:83032,MR84e:83030,Frolov:1984ra}. Thus
constructed renormalized effective action has the form 
\begin{equation}
W_{ren}\,=\,{\frac{1}{32\pi ^{2}m^{2}}}\int
d^{4}x\,g^{1/2}\sum_{n=3}^{\infty }{\frac{(n-3)!}{(m^{2})^{n-2}}}
[a_{n}(x,x^{\prime })],
\end{equation}
where $[a_{n}(x,x^{\prime })]$ is a coincidence limit of the {\it n}-th
Hadamard-Minakshisundaram-DeWitt-Seely (HMDS) coefficient. The
coefficients are local, geometrical objects constructed from curvature
tensor and its covariant derivatives of rapidly growing complexity. So far,
only coefficients for $n\leq 4$ are known, but it seems the effective action
constructed from higher order terms would be practically intractable,
especially in the attempts to calculate the stress-energy tensor.

The first nonvanishing term of the effective action constructed for the
massive scalar field with arbitrary curvature coupling satisfying 
\begin{equation}
\left( -\Box \,+\,\xi R\,+\,m^{2}\right) \phi \,=\,0,
\end{equation}
where $\xi $ is the coupling constant and $m$ is the mass of the field, form
the coincidence limit of the HMDS coefficient $\left[ a_{3}\right] ,$ is
given by \cite{Avramidi:1986mj,Avramidi:1989ik,Avramidi:1991je}:

\begin{eqnarray}
W_{ren}^{(1)}\, &=&\,\frac{1}{192\pi ^{2}m^{2}}\int d^{4}xg^{1/2}\left[ {{
\frac{1}{2}}}\left( \eta ^{2}-{{\frac{\eta }{15}}}-{{\frac{1}{315}}}\right)
R\Box R\,+\,{\frac{1}{140}}R_{pq}\Box R^{pq}\right.   \nonumber \\
&&-{\eta }^{3}R^{3}\,+\,{\frac{1}{30}\eta }RR_{pq}R^{pq}\,-\,{\frac{1}{30}
\eta }RR_{pqab}R^{pqab}  \nonumber \\
&&-{\frac{8}{945}}R_{q}^{p}R_{a}^{q}R_{p}^{a}\,+\,{\frac{2}{315}}
R^{pq}R_{ab}R_{~p~q}^{a~b}\,+\,\,{\frac{1}{1260}}R_{pq}R_{~cab}^{p}R^{qcab} 
\nonumber \\
&&+\left. {\frac{17}{7560}}{R_{ab}}^{pq}{R_{pq}}^{cd}{R_{cd}}^{ab}\,\,-\,{
\frac{1}{270}}R_{~p~q}^{a~b}R_{~c~d}^{p~q}R_{~a~b}^{c~d}\right]   \nonumber
\\
&=&
{\displaystyle{1 \over 192\pi ^{2}m^{2}}}
\sum_{i=1}^{10}\alpha _{i}W_{\left( i\right) },  \label{poteff}
\end{eqnarray}
where $\eta =\xi -1\slash6$ and $\alpha _{i\text{ }}$ are numerical
coefficients that stand in front of geometric terms in $W_{ren}^{(1)}\,.$
Differentiating functionally $W_{ren}^{\left( 1\right) }$ with respect to a
metric tensor one obtains 
\begin{eqnarray}
\,\langle T^{ab}\rangle  &=&\sum_{i=1}^{10}\alpha _{i}\tilde{T}^{\left(
i\right) ab}=
{\displaystyle{1 \over 96\pi ^{2}m^{2}g^{1/2}}}
\sum_{i=1}^{10}\alpha _{i}
{\displaystyle{\delta W_{\left( i\right) } \over \delta g_{ab}}}
\label{tens_gen} \nonumber \\
&=&T^{\left( 0\right) ab}+\eta T^{\left( 1\right) ab}+\eta ^{2}T^{\left(
2\right) ab}+\eta ^{3}T^{\left( 3\right) ab},	 \label{ten}
\end{eqnarray}
where each $\tilde{T}^{\left( i\right) ab}$ is rather complicated expression
constructed from the curvature tensor, its covariant derivatives, and
contractions. Such calculations have been undertaken in Refs. \cite
{Jirinek00prd,Jirinek01prd1}, where generic expressions for the first
nonvanishing order of the renormalized stress-energy tensor were obtained.
They generalize earlier results of Frolov and Zel'nikov for vacuum type-D\
geometries \cite{Frolov:1984ra}. One can easily extend this result
to the case of spinor and vector field as the analogous expressions differ
only by numerical coefficients $\alpha _{i}$.

It should be emphasised, however, that the above assumptions place severe
limitations on the domain of validity of the obtained approximation.
Especially, it would be meaningless, at least in this formulation, to
consider the massless limit of the approach. Therefore, the result, which
consists of approximately one hundred local geometrical terms, may be used in
any spacetime provided the temporal changes of the background geometry are
slow and the mass of the field is sufficiently large. Because of the
complexity of the thus obtained tensor it will be not presented here, and
for its full form and technical details the reader is referred to \cite
{Jirinek01prd1}.

For quantized massive scalar fields with arbitrary curvature coupling in
static and spherically-symmetric geometries there exists a different method
invented by Anderson, Hiscock, and Samuel \cite{AHS95}. Their calculations
were based on the WKB approximation of the solutions of the radial equation
and summation of thus obtained mode functions. Both methods are equivalent:
to obtain the term proportional to $m^{-2}$ one has to use 6-th order WKB,
and explicit calculations carried out for the Reissner-Nordstr\"{o}m
spacetime as well as in wormhole geometries considered in Ref.~\cite{THA97} yielded
identical results. Moreover, detailed numerical analyses carried out in Ref. 
\cite{AHS95} and briefly reported in \cite{THA00} show that for $m M\gtrsim 2$ 
($M$ being the black hole mass),
the accuracy of the Schwinger-DeWitt approximation in the Reissner-Nordstr\"{o}m
geometry is quite good (1\% or better).

Geometry of the Reissner-Nordstr\"{o}m black hole is
singular as $r \to 0$ and there is a natural desire to
construct its regular generalizations. It is expected that a
good candidate for the source term of the Einstein equations
is the (classical) stress-energy tensor of nonlinear
electrodynamics. Moreover, renewal of interests in nonlinear
electrodynamics that is seen recently is motivated by the
observation that such theories arose as limiting cases of
certain formulations of string theory. Unfortunately, the no
go theorem proved in Ref. \cite{Bronnikov:1979ex} clearly
shows that for  Lagrangians ${\cal L}(F),$ $(F=F_{ab}F^{ab}),$ with
the Maxwell weak-field limit there are no 
spherically-symmetric static black hole solutions with a regular center.

Recently, employing the Schwinger-DeWitt approximation we have constructed
the renormalized stress-energy tensor of the quantized conformally
coupled massive scalar field in the spacetime of the electrically charged regular
black hole being an exact solution of the equations of the nonlinear
electrodynamics and the Einstein field equations. Such a solution has been
proposed by Ay\'{o}n-Beato and Garc\'{i}a (ABG) in Ref.~\cite{Ayon-Beato:1999rg}. It
should be noted, however, the ABG line element is not a solution of the standard
nonlinear electrodynamics and the effective geometry (i.e. the geometry
seen by photons of the nonlinear theory) is singular. Fortunately, the ABG
solution may be reinterpreted as describing a magnetically charged regular
solution of the coupled equations of standard nonlinear electrodynamics and
gravitation with much more regular behaviour of the effective geometry \cite
{Bronnikov:2000vy}. Moreover, it has been shown recently that it is possible to
combine electric and magnetic line elements to obtain a regular electric
solution with a magnetic core \cite{Burinsky:2002pz}. 

For small and intermediate charges as well as at large distances, the
geometry of the ABG black hole resembles that of RN; noticeable differences
appear near the extremality limits. It is interesting therefore to analyse
how the similarities in the metric structure of the ABG and RN black holes
are reflected in the structure of the stress-energy tensors. As for the conformally
coupled massive scalar field the linear, quadratic, and cubic terms in $\eta $ 
in the stress-energy tensor (\ref{tens_gen}) are absent, it is anticipated 
that such similarities do
occur. Explicit calculations carried out in \cite{Jirinek01prd1} confirmed
this hypothesis and showed that for small values of $q$ the appropriate
tensors are practically indistinguishable, and, as expected, important
differences appear near and at the extremality limit.

In this paper we shall extend the results of Ref. \cite{Jirinek01prd1} and
investigate a much more complicated case of arbitrary $\eta $ with special
emphasis put on the minimal coupling. Although complexity of the approximate
stress-energy tensor constructed for the ABG line element practically
prevents its direct examination, it is possible to extract interesting
information expanding $\langle T_{a}^{b}\rangle $ into a power series and
retaining a few leading terms. We shall show that such analyses could be
carried out for small charges, large distances, and in the vicinity of the
event horizon of the extremal ABG black hole. Moreover, on general grounds
one may easily estimate the role played by $\tilde{T}^{\left( 1\right) ab}$
and $\tilde{T}^{\left( 3\right) ab}.$ To gain a deeper understanding of the
problem, however, we employ numerical calculations.

The paper is organized as follows. In Sec. II the essentials of the ABG
black hole geometry that are neccessary in further development are briefly
described. In Sec. III certain features of the approximate stress-energy
tensor of the massive scalar field in the ABG geometry are discussed and
compared with the appropriate tensors constructed in the RN geometry. The
results of our numerical analyses are presented in Sec. IV in which we
discuss behaviour of the components of $\langle T_{a}^{b}\rangle $\ in some
detail and present it graphically.

\section{Geometry}

\label{geo}

The general Reissner-Nordstr\"{o}m solution of the Einstein-Maxwell
equations describing a static and spherically symmetric black hole of mass $M,$
electric charge $e,$ and magnetic monopole charge $e_{m}$ has a remarkably
simple form 
\begin{equation}
ds^{2}=-A\left( r\right) dt^{2}+A^{-1}\left( r\right) dr^{2}+r^{2}\left(
d\theta ^{2}+\sin ^{2}\theta d\phi ^{2}\right) ,  \label{genline}
\end{equation}
with 
\begin{equation}
A\left( r\right) =1-{{\frac{2M}{r}}}\,+\,{{\frac{e^{2}+e_{m}^{2}}{r^{2}}}}.
\label{rnA}
\end{equation}
Since the charges enter (\ref{rnA}) as a sum of squares the metric structure
remains unchanged under the changes of charges leaving $
Q^{2}=e^{2}+e_{m}^{2} $ constant. For $Q^{2}/M^{2}<1,$ the equation $A\left(
r\right) =0$ has two positive roots 
\begin{equation}
r_{\pm }=M\pm \sqrt{M^{2}-Q^{2}},
\end{equation}
and the larger one, $r_{+},$ determines the location of the event horizon,
whereas the smaller, $r_{-}$, gives the position of the inner horizon. In
the limit $Q^{2}=M^{2\text{ }}$ horizons merge at $r=M$ and the black hole
degenerates to the extremal one. At $r=0$ the Reissner-Nordstr\"{o}m solution
has an irremovable curvature singularity, which, although hidden to an
external observer for $Q^{2}/M^{2}\leq 1,$ remains a somewhat unwanted feature
of the solution. The solution for $Q^{2}>M^{2\text{ }}$ is clearly
unphysical.

At least for purely electrical solutions the nonlinear electrodynamics does
not remedy the situation. Indeed, consider ralization of the nonlinear 
electrodynamics with
the action functional of the form 
\begin{equation}
S=
{\displaystyle{1 \over 16\pi }}
\int d^{4}x\sqrt{-g}\left[ R-{\cal L}\left( F\right) \right] ,
\label{standard}
\end{equation}
where $R$ is a curvature scalar, $F=F_{ab}F^{ab},$ and ${\cal L}\left(
F\right) $ is an arbitrary function with Maxwell asymptotic in a weak field
limit, i. e. ${\cal L}\left( F\right) \rightarrow F$ \ and $\frac{d}{dF}
{\cal L}\left( F\right) \rightarrow 1$ \ \ as \ $F\rightarrow 0$. According
to a well known theorem \cite{Bronnikov:1979ex,Bronnikov:2000vy}, there are
no regular, static, and spherically symmetric solutions of general
relativity coupled to nonlinear electrodynamics describing a black hole with
nonzero electric charge. However, as has been explicitly demonstrated
recently by Bronnikov, adopted hypotheses leave room for appropriate
regular solutions with a nonzero {\em magnetic} charge. In this regard it is
interesting to note that within a different formulation of the nonlinear
electrodynamics proposed in \cite{Salazar+} (a ${\cal P}$ framework
according to the nomenclature of Refs. \cite{Bronnikov:2000yz,Bronnikov:2000vy})
obtained from the standard one (the ${\cal F}$ formulation) (\ref{standard})
by means of a Legendre transformation, Ay\'{o}n-Beato and Garc\'{i}a
constructed a regular black hole solution with a nonzero electric charge and
mass~\cite{Ayon-Beato:1999rg}.  Their solution has the simple form 
(\ref{genline}), where 
\begin{equation}
A\left( r\right) =1-{{\frac{2M}{r}}}\left( 1-\tanh {{\frac{e^{2}}{2Mr}}}
\right) .  \label{metpot}
\end{equation}
Bronnikov also demonstrated that any spherically symmetric solution
constructed within the {\it \ }${\cal F}${\it \ }framework has its
counterpart with the same metric tensor constructed within the ${\cal P}$
framework, and therefore the electric solution (\ref{genline}) has the
magnetic companion with $e$ replaced by the magnetic charge $e_{m}.$
Moreover, Burinskii and Hildebrandt have proposed recently a regular hybrid model
in which electrically and magnetically charged solutions were combined in
such a way that the electric field does not extend to the center of the black hole
\cite{Burinsky:2002pz}.

Although the magnetic and electric ABG solutions are precisely of the type 
(\ref{genline}) with (\ref{metpot}), the geometries seen by photons of the
nonlinear theory are different. It is because photons of the nonlinear
theory move along null geodesics of the effective metric, and the latter is
singular for the electric solution \cite{nov1,nov2,nov3}. It should be noted,
however, that otherwise the physical geometry is regular and well behaving.

Since geometries outside the event horizon are described by
the same line element and since we are going to consider
neutral scalar field only, our results will hold for any
particular realization of the ABG black hole as the only
concern here is the metric structure of the spacetime.
Therefore in what follows we denote  both electric and
magnetic charge by $e.$

Inspection of the metric potentials reveals interesting features of the
ABG solution: its curvature invariants are finite as $r$ $\rightarrow 0$ and
at large $r$ it approaches the Reissner-Nordstr\"{o}m solution. Moreover,
for small and intermediate values of the ratio $e^{2}/M^{2},$ the radial
coordinate of the event horizon, $r_{+},$ is close to the event horizon of
the RN black hole; significant differences occur near the extremality limit.

It has been shown in \cite{Jirinek01prd1} that the location of the horizons
of the ABG black hole may be expressed in terms of the Lambert function $W$ 
\cite{Lambert}$.$ Indeed, making use of the substitution $r=Mx$, 
$e^{2}=q^{2}M^{2},$ it could be demonstrated that the location of the horizons is
given by the real branches of the Lambert functions: 
\begin{equation}
x_{\pm }\,=\,-{\frac{4q^{2}}{4W(\varepsilon ,-{q^{2}/4\,}\exp
(q^{2}/4))-q^{2}}},
\end{equation}
where $\varepsilon $ is $0$ for the event horizon, $x_{+\text{ }},$ and $-1$
for the inner one, $x_{-}.$ The functions $W(0,s)$ and $W(-1,s)$ are the
only real branches of the Lambert function with the branch point at $s=-{\rm 
e}^{-1},$ where {\rm e} is the base of natural logarithms and 
\begin{equation}
W(0,-{\rm e}^{-1})=W(-1,-{\rm e}^{-1})=-1.
\end{equation}
Consequently, for 
\begin{equation}
q=\,2\sqrt{w}
\end{equation}
the horizons merge at 
\begin{equation}
x_{extr}\,=\,{\frac{4w}{1+w}},
\end{equation}
where $w=W\left( 0,{\rm e}^{-1}\right) =0.2785.$ Numerically, one has $
x_{extr}\,=\,0.871$ for $\left| e\right| \slash M=\,1.055.$

For small values of $q,$ the ABG line element resembles that of RN. Indeed,
expanding the function $A\left( x\right) $ one has

\begin{equation}
A\left( x\right) =1-\frac{2}{x}+\frac{q^{2}}{x^{2}}+{\cal O}\left(
q^{4}\right) .
\end{equation}
In spite of similarities between ABG and RN geometries for $q\ll 1,$ there
are substantial differences in the extremality limit: the extremal RN
line element is described by (\ref{genline}) with 
\begin{equation}
A(x)\,=\,\left( 1-{\frac{1}{x}}\right) ^{2},  \label{extr1}
\end{equation}
whereas near the event horizon of the extremal ABG black hole the function $A(x)$
has the following expansion 
\begin{equation}
A(x)\,=\,(x-x_{extr})^{4}\,+\,{\frac{(1+w)^{3}}{32w^{2}}}(x-x_{extr})^{5}\,+
\,O(x-x_{extr})^{6}.  \label{extr2}
\end{equation}
The\ form of the above expansion indicates that the proper length between
the event horizon and any point located at $r>r_{extr}$ is infinite.

\section{The renormalized stress-energy tensor}

\label{tens}

Structure of terms in the effective potential (\ref{poteff})
and in Eq.~(\ref{tens_gen}) indicates that for the
conformally coupled massive scalar field, i.e. for $\eta
=0,$ there are substantial simplifications of the
renormalized stress-energy tensor as in this very case the
third, fourth, and fifth term in (\ref{poteff}) do not
contribute to the final result. Moreover, from
(\ref{poteff}) it is evident that in $R=0$ geometries and
for arbitrary curvature coupling the functional derivative of
the first and the third terms in (\ref{poteff}) with respect
to the metric tensor vanishes and, therefore, the
approximate renormalized stress-energy tensor of the massive field in RN spacetime
has a general form
\begin{equation}
T_{a}^{b}=C_{a}^{b}+\eta D_{a}^{b}.  \label{rn1}
\end{equation}
On the other hand, however, for the ABG metric one has

\begin{equation}
R={{\frac{{q}^{4}}{M^{2}x^{5}}}}\tanh \,\left( {\frac{{q}^{2}}{2x}}\right) 
\left[ 1-\tanh ^{2}\left( {\frac{{\ q}^{2}}{2x}}\right) \right] ,
\end{equation}
which for small $q$ is ${\cal O}\left( q^{6}\right) .$ Detailed calculations
carried out for $q\ll 1$ show that for the ABG solution neither $\tilde{T}
_{a}^{\left( 1\right) b}$ nor $\tilde{T}_{a}^{\left( 3\right) b}$ contribute
importantly to the result.

Because of the similarities of the metric structure of the
RN and ABG solutions, the overall behaviour of the
renormalized stress-tensors for conformal coupling should be
qualitatively similar and comparable in magnitude at least
for small and intermediate values of $q.$ However, the
differences between the line element (\ref{extr1}) and the
expansion (\ref{extr2}), give strong evidence for
differences in the appropriate components of the approximate
stress-energy tensors. For the conformally coupled
massive scalar field this statement has been confirmed by extensive
numerical calculations reported in Ref.~\cite
{Jirinek01prd1}. Now we shall analyse the case of an arbitrary coupling
with curvature. First, observe that contribution of
$\eta^{2}{\tilde T}^{\left( 1\right) ab}$ 
and 
$ \eta ^{3}{\tilde T}^{\left(3\right) ab}$ 
to the stress-energy tensor could be made
arbitrarily great by a suitable choice of the conformal
coupling. It should be noted, however, that such great
values of $\eta $ are clearly unphysical and should be
rejected. Therefore, as the particular case of the conformal
coupling has been considered earlier, here we shall confine
ourselves mostly to the important and physically interesting case of
minimal coupling $\eta =-1\slash6.$

In order to construct $C_{a}^{b}$ and $D_{a}^{b}$ one has to either
calculate the curvature tensor and its covariant derivatives to required
order for the line element (\ref{genline})  with (\ref{rnA}) and employ
the general formulas 
presented in \cite{Jirinek00prd} or to make use of the method proposed by
Anderson, Hiscock and Samuel \cite{AHS95}. Both approaches yield the same
result, which reads:

\begin{equation}
C_{t}^{t}=\,{\frac{810\,{x}^{4}{q}^{2}-855\,{x}^{4}-202\,{x}^{2}{q}
^{4}+1878\,{x}^{3}-1152\,{x}^{3}{q}^{2}-2307\,{x}^{2}{q}^{2}+3084\,x{q}
^{4}-1248\,{q}^{6}}{30240{M}^{6}{\pi }^{2}{m}^{2}{x}^{12}},}  \label{rn1a}
\end{equation}

\begin{equation}
D_{t}^{t}={\frac{-1008\,{x}^{3}{q}^{2}+819\,{q}^{6}+2604\,{x}^{2}{q}
^{2}+728\,{x}^{2}{q}^{4}-2712\,x{q}^{4}+360\,{x}^{4}-792\,{x}^{3}}{720{M}^{6}
{\pi }^{2}{m}^{2}{x}^{12}},}
\end{equation}

\begin{equation}
C_{t}^{t}={\frac{842\,{x}^{2}{q}^{4}+444\,{q}^{6}+162\,{x}^{4}{q}^{2}-462\,{x
}^{3}-1488\,{x}^{3}{q}^{2}-1932\,x{q}^{4}+315\,{x}^{4}+2127\,{x}^{2}{q}^{2}}{
30240{M}^{6}{\pi }^{2}{m}^{2}{x}^{12}},}
\end{equation}

\begin{equation}
D_{r}^{r}={\frac{504\,x{q}^{4}-208\,{x}^{2}{q}^{4}-588\,{
x}^{2}{q}^{2}+336\,{x}^{3}{q}^{2}-117\,{q}^{6}+216\,{x}^{3}-144\,{x}^{4}}{720
{M}^{6}{\pi }^{2}{m}^{2}{x}^{12}},}
\end{equation}

\begin{equation}
C_{\theta }^{\theta }={\frac{2202\,{x}^{3}-486\,{x}^{4}{q}^{2}-945\,{x}^{4}-3044\,{x}
^{2}{q}^{4}+4884\,{x}^{3}{q}^{2}-9909\,{x}^{2}{q}^{2}+10356\,x{
q}^{4}-3066\,{q}^{6}}{30240{M}^{6}{\pi }^{2}{m}^{2}{x}^{12}}}
\end{equation}
and

\begin{equation}
D_{\theta }^{\theta }=\,{\frac{-1176\,{x}^{3}{q}^{2}+1053\,{q}^{6}+3276\,{x}
^{2}{q}^{2}+832\,{x}^{2}{q}^{4}-3408\,x{q}^{4}+432\,{x}^{4}-1008\,{x}^{3}}{
720{M}^{6}{\pi }^{2}{m}^{2}{x}^{12}}.}  \label{rnlast}
\end{equation}
The analogous tensor in the ABG geometry is more complicated, and besides the
linear terms it contains also terms which are quadratic and cubic in $\eta .$
Each component of the stress-energy tensor consists of more than three
hundred terms and has a general form 
\begin{equation}
{\displaystyle{1 \over 96\pi ^{2}m^{2}M^{6}}}
\left( 1-\tanh \frac{q^{2}}{2x}\right) \sum_{i,j,k} \alpha _{ijk}\frac{q^{2i}}{x^{j}}
\tanh ^{k}\frac{q^{2}}{2x},
\label{term}
\end{equation}
where $\alpha _{ijk}$ for a given $\eta $ are numerical coefficients 
$(0\leq i\leq 6,$ $8\leq j\leq 15$, $0\leq k\leq 8),$ and for obvious reasons
will not be presented here. 

Since the form of the Eq.~(\ref{term}) differs
considerably from the stress-energy tensor given by
(\ref{rn1}) with (\ref{rn1a}--\ref{rnlast}), it
could be expected that the runs of both tensors have 
nothing in common. On the other hand  similarities
in the metric structure  of the RN and ABG black
holes discussed earlier strongly suggest the
opposite. 
Unfortunately, the complexity of the renormalized stress-energy tensor of
the massive scalar field in the ABG geometry practically
invalidates direct analytical treatment. One can, however,
obtain interesting and important information analysing
certain limiting cases. Indeed, expanding the stress-energy
tensor for $q\ll 1$ one has

\begin{equation}
\,\langle T_{a}^{b}\rangle ^{ABG}=T_{a}^{b}+\Delta _{a}^{\left( 1\right) b}+
{\cal O}\left( q^{8}\right),  \label{male_q}
\end{equation}
where $T_{a}^{b}$ is given by (\ref{rn1}--\ref{rnlast}) and

%skladowe tep w RN

\begin{eqnarray}
96\pi ^{2}m^{2}\Delta _{t}^{(1)}{}^{t} &=&-\frac{q^{6}\,\left(
12049-11416\,x+2660\,x^{2}\right) }{420\,M^{6}\,x^{12}}+\frac{
2\,q^{6}\,\left( 617-596\,x+140\,x^{2}\right) \,\eta }{5\,M^{6}\,x^{12}} 
\nonumber \\
&&-\frac{12\,q^{6}\,\left( 459-366\,x+70\,x^{2}\right) \,{\eta }^{2}}{
M^{6}\,x^{12}},  \label{exp1}
\end{eqnarray}

\begin{eqnarray}
96\pi ^{2}m^{2}\Delta _{r}^{(1)}{}^{r} &=&-\frac{q^{6}\,\left(
1555-1064\,x+140\,x^{2}\right) }{420\,M^{6}\,x^{12}}-\frac{2\,q^{6}\,\left(
121-140\,x+40\,x^{2}\right) \,\eta }{5\,M^{6}\,x^{12}}  \nonumber \\
&&+\frac{12\,q^{6}\,\left( 81-84\,x+20\,x^{2}\right) \,{\eta }^{2}}{
M^{6}\,x^{12}},
\end{eqnarray}

\begin{eqnarray}
96\pi ^{2}m^{2}\Delta _{\theta }^{(1)}{}^{\theta } &=&\frac{q^{6}\,\left(
5249-3528\,x+560\,x^{2}\right) }{420\,M^{6}\,x^{12}}+\frac{2\,q^{6}\,\left(
779-720\,x+160\,x^{2}\right) \,\eta }{5\,M^{6}\,x^{12}}  \nonumber \\
&&-\frac{12\,q^{6}\,\left( 540-423\,x+80\,x^{2}\right) \,{\eta }^{2}}{
M^{6}\,x^{12}}.  \label{exp3}
\end{eqnarray}
Inspection of expansions (\ref{exp1}--\ref{exp3}), which are valid for any 
$\eta $ and $x,$ shows that in this order the term $\tilde{T}_{\left(
3\right) }^{ab}$ is absent in the final result. It is because $\tilde{T}
_{a}^{\left( 3\right) b}\sim {\cal O}\left( q^{12}\right).$ The second term  
$\tilde{T}_{a}^{\left( 1\right) b},$
which vanishes in the RN geometry is now 
${\cal O}\left( q^{6}\right).$

A similar expansion valid for large values of radial coordinate has the
following form

\begin{equation}
\langle T_{a}^{b}\rangle ^{ABG}=T_{a}^{b}+\Delta _{a}^{\left( 1\right)
b}+\Delta _{a}^{\left( 2\right) b}+{\cal O}\left( x^{-13}\right) ,
\end{equation}
where $T_{a}^{b}$ is given by (\ref{rn1}--\ref{rnlast}), $\Delta
_{a}^{\left( 1\right) b}$ by (\ref{exp1}--\ref{exp3}), and

\begin{equation}
96\pi ^{2}m^{2}\Delta _{t}^{(2)}{}^{t}=\frac{q^{8}\,\left(
-2574+1015\,q^{2}\right) }{210\,M^{6}\,x^{12}}-\frac{14\,q^{8}\,\left(
-39+20\,q^{2}\right) \,\eta }{5\,M^{6}\,x^{12}}+\frac{105\,q^{8}\,\left(
-27+8\,q^{2}\right) \,{\eta }^{2}}{M^{6}\,x^{12}},  \label{p1}
\end{equation}

\begin{equation}
96\pi ^{2}m^{2}\Delta _{r}^{(2)}{}^{r}=\frac{q^{8}\,\left(
-738+35\,q^{2}\right) }{378\,M^{6}\,x^{12}}+\frac{4\,q^{8}\,\left(
-51+28\,q^{2}\right) \,\eta }{9\,M^{6}\,x^{12}}-\frac{5\,q^{8}\,\left(
-333+112\,q^{2}\right) \,{\eta }^{2}}{3\,M^{6}\,x^{12}},
\end{equation}

\begin{equation}
96\pi ^{2}m^{2}\Delta _{\theta }^{(2)}{}^{\theta }=\frac{q^{8}\,\left(
12456-875\,q^{2}\right) }{1890\,M^{6}\,x^{12}}-\frac{112\,q^{8}\,\left(
-12+5\,q^{2}\right) \,\eta }{9\,M^{6}\,x^{12}}+\frac{5\,q^{8}\,\left(
-1989+560\,q^{2}\right) \,{\eta }^{2}}{3\,M^{6}\,x^{12}}.  \label{p3}
\end{equation}
Note that in Eqs. (\ref{p1}--\ref{p3}) there is no contribution from the
term $\tilde{T}_{a}^{\left( 3\right) b}$ and, consequently, the result which
is valid for any combination of couplings and charges is independent of $\eta ^{3}.$

In the vicinity of the event horizon of the extremal black hole the
renormalized stress-energy tensor has the following expansion 
\begin{equation}
96\pi ^{2}m^{2}\langle T_{a}^{b}\rangle ^{ABG}=t_{a}^{\left( 1\right)
b}+t_{a}^{\left( 2\right) b}\left( x-x_{extr}\right) +{\cal O}\left(
x-x_{extr}\right) ^{2},  \label{q_extr}
\end{equation}
\begin{eqnarray}
t_{t}^{\left( 1\right) t}=t_{r}^{\left( 1\right) r}= &&-{\frac{1}{4096}}\,{
\frac{\left( 1+w\right) ^{6}\left( 2+w\right) \left( w-1\right) ^{2}{\eta }
^{3}}{{w}^{6}{M}^{6}}}-{\frac{1}{245760}}\,{\frac{\left( 1+w\right)
^{6}\left( 4+w+{w}^{3}+2\,{w}^{2}\right) \eta }{{w}^{6}{M}^{6}}}  \nonumber
\\
&&+{\frac{1}{1935360}}\,{\frac{\left( 1+w\right) ^{6}\left( {w}^{3}+3\,{w}
^{2}+3\,w+5\right) }{{w}^{6}{M}^{6}}},  \label{eq1}
\end{eqnarray}

\begin{eqnarray}
t_{t}^{\left( 2\right) t}=t_{r}^{\left( 2\right) r}= &&{\frac{3}{16384}}\,{
\frac{\left( 1+w\right) ^{7}\left( w+3\right) \left( {w}^{2}-2\,w+1\right) \,
{\eta }^{3}}{{w}^{7}{M}^{6}}}
-{\frac{1}{2580480}}\,{\frac{\left( 1+w\right) ^{7}\left( w+3\right)
\left( {w}^{2}+3\right) }{{w}^{7}{M}^{6}}}\nonumber \\
&&+{\frac{1}{983040}}\,{\frac{\left( 1+w\right)
^{7}\left( w+3\right) \left( 3\,{w}^{2}-2\,w+7\right) \eta }{{w}^{7}{M}^{6}}},
\label{eq2}
\end{eqnarray}
\begin{eqnarray}
t_{\theta }^{\left( 1\right) \theta } &=&t_{\phi }^{\left( 1\right) \phi }={
\frac{1}{8192}}\,{\frac{\left( 1+w\right) ^{6}\left( w+5\right) \left(
w-1\right) ^{2}{\eta }^{3}}{{w}^{6}{M}^{6}}}+{\frac{1}{491520}}\,{\frac{
\left( 1+w\right) ^{6}\left( -w+13+3\,{w}^{2}+{w}^{3}\right) \eta }{{w}^{6}{M
}^{6}}}  \nonumber \\
&&-{\frac{1}{3870720}}\,{\frac{\left( 1+w\right) ^{6}\left( 3\,w+17+3\,{w}
^{2}+{w}^{3}\right) }{{w}^{6}{M}^{6}}}  \label{eq3}
\end{eqnarray}
and 
\begin{eqnarray}
t_{\theta }^{\left( 2\right) \theta } &=&t_{\phi }^{\left( 2\right) \phi }=-{
\frac{3}{32768}}\,{\frac{\left( 1+w\right) ^{7}\left( -25\,w+34-23\,{w}
^{2}+9\,{w}^{3}+5\,{w}^{4}\right) {\eta }^{3}}{{M}^{6}{w}^{7}}}  \nonumber \\
&&-{\frac{1}{16384}}\,{\frac{\left( 1+w\right) ^{7}\left( -10\,w-8+{w}
^{2}+4\,{w}^{3}+{w}^{4}\right) {\eta }^{2}}{{M}^{6}{w}^{7}}}  \nonumber \\
&&+{\frac{1}{1966080}}\,{\frac{\left( 1+w\right) ^{7}\left( -185\,w-190-111\,
{w}^{2}+5\,{w}^{4}-15\,{w}^{3}\right) \,\eta }{{M}^{6}{w}^{7}}.}  \label{eq4}
\end{eqnarray}
Since $\tilde{T}_{a}^{\left( 1\right) b}$ vanishes in the limit \ $
x\rightarrow x_{extr}$ there are no terms proportional to $\eta ^{2}$ in 
(\ref{eq1}--\ref{eq3}).

The components of the stress-energy tensor of the massive scalar field are
regular functions of the radial coordinate and are finite on the event
horizon. Moreover, it could be demonstrated that the difference between
radial and time components of $\langle T_{a}^{b}\rangle ^{ABG}$ factorizes:

\begin{equation}
\langle T_{t}^{t}\rangle ^{ABG}-\langle T_{r}^{r}\rangle ^{ABG}=\left[ 1-
{\displaystyle{2M \over r}}
\left( 1-\tanh 
{\displaystyle{e^{2} \over 2Mr}}
\right) \right] F\left( r\right) ,
\end{equation}
where $F\left( r\right) $ is a regular function and, consequently, the
stress-energy tensor is finite in a freely falling frame. To demonstrate
this let us consider a slightly more general line element 
\begin{equation}
ds^{2}=-f\left( x^{1}\right) \left( dx^{0}\right) ^{2}+g\left( x^{1}\right)
\left( dx^{1}\right) ^{2}+\left( x^{1}\right) ^{2}d\Omega ^{2}\text{.}
\end{equation}
For a radial motion the vectors of the frame are the four-velocity $
e_{0}^{\alpha }=u^{\alpha }$ and a unit length spacelike vector $
e_{1}^{\alpha }=n^{\alpha }$. Then, using the geodesic equation, one finds 
\begin{equation}
e_{\left( 0\right) }^{a}=u^{a}=\left( \frac{\gamma }{f},-\sqrt{\left( \frac{
\gamma ^{2}}{f}-1\right) \frac{1}{g}},0,0\right) 
\end{equation}
and 
\begin{equation}
e_{\left( 1\right) }^{a}=n^{a}=\left( -\frac{1}{f}\sqrt{\gamma ^{2}-f},\frac{
\gamma }{\sqrt{fg}},0,0\right) ,
\end{equation}
where $\gamma $ is the energy per unit mass along the geodesic. A simple
calculation shows that the components $T_{(0)(0)}$, $T_{(0)(1)}$ and $
T_{(1)(1)\text{ }\,}$in a freely falling frame (independent of the function $
g\left( x^{1}\right) $) are 
\begin{equation}
T_{(0)(0)}=\frac{\gamma ^{2}(T_{1}^{1}-T_{0}^{0})}{f}-T_{1}^{1}\text{,}
\label{fram1}
\end{equation}
\begin{equation}
T_{(1)(1)}=\frac{\gamma ^{2}(T_{1}^{1}-T_{0}^{0})}{f}+T_{1}^{1}\text{,}
\end{equation}
\begin{equation}
T_{(0)(1)}=-\frac{\gamma \sqrt{\gamma ^{2}-f}(T_{1}^{1}-T_{0}^{0})}{f}\text{.
}  \label{fram3}
\end{equation}
One concludes, therefore, that if  all components of $T_{a}^{b}$ and 
\begin{equation}
{(T_{1}^{1}-T_{0}^{0})\over f}
\label{regularity}
\end{equation}
are finite on the horizon, the stress-energy tensor in a
freely falling frame is finite as well.	 It is expected that
constructed formulas satisfactorily approximate the exact
stress-energy tensor. On the other hand however, to
establish the regularity of the  exact $\langle
T^{b}_{a}\rangle^{ABG}$ in a freely falling frame one has to
explicitly demonstrate that (\ref{regularity}) is
satisfied.

%%%%%
%%%%%
%%%%%
%%%%%
\section{Numerical results}

\label{num}

Considerations of the previous section are limited to
analytically tractable special cases. However, to gain
insight into the overall behavior of the stress-energy
tensor for any combination of couplings and charges one has
to refer to numerical calculations --- our complete but
rather complicated analytical formulas are, unfortunately,
not of much  help in this regard. Below we describe the main
features of the constructed tensors and present them
graphically, fixing our attention on the physically
interesting case of minimal coupling. Related plots showing
the radial dependence of $C_{a}^{b}$ and $T^{(0)b}_{a}$ can
be found in Ref.~\cite{Jirinek01prd1}, where the discussion
of the stress-energy tensor of the massive scalar field with
conformal coupling with curvature has been carried out.

The case of arbitrary $\eta$ is much more complicated as the
tensor $\langle T_{a}^{b}\rangle^{ABG},$ given by
(\ref{ten}), is modified by the presence of additional
terms. However, numerical analysis performed for $\eta =
-1/6$ reveals that the contribution of
$\eta^{3}T^{(3)b}_{a}$ is neglegible for $q$ even as large
as $1.0$. Moreover, a closer examination indicates that as
one approaches the extremality limit, the magnitude of this
very term becomes comparable with $\eta^{2}T^{(2)b}_{a}$ and
$\eta T^{(1)b}_{a}$ only in the closest vicinity of the event
horizon. Our calculations also suggest that irrespective of
the exact value of $q,$ each component of $T^{(1)b}_{a}$ is
a monotonic function of the radial coordinate in a large
neighbourhood of the event horizon, if not in the whole
range $r>r_{+}.$ For the extremal case such a behaviour is
illustrated in Figs.~1--3, where the time, radial, and
angular components of $T^{(1)b}_{a}$ are displayed. For
comparison and completeness we also present the run of
$T^{(0)b}_{a},$  $C_{a}^{b},$ and $D_{a}^{b},$ where the
latter two tensors have been calculated for the RN line
element with the aid of Eqs.~(\ref{rn1a}--\ref{rnlast}). It
should be noted that even in the extremal case, when the
differences between two considered geometries are most
prominent, the tensors $T^{(1)b}_{a}$ and $D_{a}^{b}$ behave
in a similar manner though differing noticeably in
magnitude. For small values of $q,$ the curves constructed
for both types of black hole are almost indistinguishable,
which could be easily established from Eq.~(\ref{male_q}).

On the other hand however, the components of $T^{(2)b}_{a}$
exhibit quite different but very regular behavior, which
will be described in some detail. For $q\ll 1$ it can be, of
course, inferred from the approximate formulas
(\ref{male_q}--\ref {exp3}); greater values of $q$ require
numerical examination. Specifically, for small charges each
component of the considered tensor is negative at $r=r_{+},$
and before approaching zero as $r\rightarrow \infty ,$ it
changes sign once for the radial component and twice for
time and angular components. With the increase of $q,$ all
components become strongly negative at $r_{+},$ with their
zeros shifted towards larger values of the radial
coordinate. Near the extremality limit, the tensor
$T^{(2)b}_{a}$ is very sensitive to changes of $q,$ which is
illustrated in Figs. 4--6. If the value of ratio $\left|
e\right| /M$ slightly exceeds 1.0, for all components there
occurs a negative minimum not far away from the event
horizon. For the extremal ABG black hole $T^{(2)b}_{a}$
vanishes at $r_{+},$ as is clearly seen from Eqs.
(\ref{q_extr}--\ref{eq3}), and exhibits oscillatory-like
behavior with a rapidly decreased amplitude and increased
intervals between zeros. The angular component attains a
very distinct maximum near the event horizon whereas for
time and radial component there are minima close to $r_{+}$.

The competition of the terms described above results in the
overall behaviour of the stress-energy tensor of minimally
coupled massive fields in the geometry of the ABG black
hole. Numerical calculations indicate that  for small and
intermediate values of $q,$ up to about $0.8,$ $\langle
T^{b}_{a}\rangle^{ABG}$ still resembles that  evaluated for
conformal coupling. For larger values of $q,$ however, the
change of curvature coupling leads to a considerable
modification of the results as both magnitudes and radial
variations become significantly different. This is
illustrated in Figs. 7--9, where values of $q$ at and near
the extremality limit were chosen. The oscillatory-like
behaviour of $T^{(2)b}_{a}$ for $q$ close to $q_{extr}$  is
refected  by the presence of local extrema	visible in the
stress-energy tensor, and, for the angular component, an
inflection point. It is in a sharp contrast with the almost
monotonic behavior of  the stress-energy tensor of
conformally coupled massive fields depicted in Fig. 7--9 by
the dashed lines. Moreover, it should be noted that for
$\eta = -1/6,$ when the extremality limit is approached, there
are substantial changes of values of $\langle
T^{b}_{a}\rangle^{ABG}$ at the event horizon as well as in
the behavior of the stress-energy tensor in a narrow strip
near  $r_{extr}$. There appear also certain new features for
the time and radial components which are not present in the
case of the conformal coupling. Indeed, for $0.987\lesssim
q\lesssim 1.032,$ the energy density $\rho = -\langle
T^{t}_{t}\rangle^{ABG}$ is positive at the event horizon
whereas for the same values of $q,$ the horizon value of
the radial pressure $p_{r} = \langle T_{r}^{r}\rangle^{ABG}$
is negative. It should be noted in this regard that for
$\eta=0$ the radial component of the stress-energy tensor is
always positive there. On the other hand, the angular
pressure is positive on the event horizon for $\left|
e\right| /M \lesssim 0.922$ and negative for larger values
that is very similar to the previously studied case of
conformal coupling.

\section{Concluding remarks}

\label{conclud}

In this work our goal was to construct the renormalized
stress-energy tensor of the quantized massive field in the
spacetime of a nonlinear black hole and to investigate how
the choice of the curvature coupling affects the results. A
regular electrically charged solution of this type has been
recently proposed by Ay\'{o}n-Beato and Garc\'{i}a in ${\cal
P}$ formulation of nonlinear electrodynamics and
reinterpreted by Bronnikov as a regular magnetically charged
solution of the standard ${\cal F}$ formulation. For small
and intermediate values of the ratio $\left| e\right| /M$
the metric structure of the nonlinear solution closely
resembles that of Reissner and Nordstr\"{o}m and the
similarities in the line elements are reflected in the
behavior of the stress-energy tensor of the conformally
coupled massive scalar fields; notable differences appear
near and at the extremality limit.

As the general $\langle T_{a}^{b}\rangle^{ABG} $ contains
terms that are quadratic and cubic in $\eta$, the case of
arbitrary coupling is more complicated. Again, for small and
intermediate values of $q$ there is a similarity between the
$\langle T^{b}_{a}\rangle$ evaluated for the minimally
coupled scalar field in the ABG geometry and its RN
counterpart. Modifications for $q$ slightly exceeding 0.8,
which are mainly due to the term $T^{(2)b}_{a},$ are
noticeable although there are still some similarities.
Comparison of $\langle T_{a}^{b}\rangle^{ABG}$ for charges
close to $q_{extr}$ indicates that the radial dependences
for both couplings are completely different. Indeed, the
behaviour of the stress-energy tensor for the minimal
coupling is far more complicated as compared with its almost
monotonic radial dependence  for $\eta = 0$. Moreover, the
former is even more sensitive to changes of q, especially
near the extremality limit.  In addition, new features occur
as, for example, positivity of the energy density at the
event horizon for certain values of $q.$

It should be emphasized, that being local the
Schwinger-DeWitt approximation does not describe particle
creation, and, therefore, it must not be applied in strong
or rapidly varying gravitational fields. However, it is
expected that for sufficiently massive fields the method
provides a good approximation  of the exact renormalized
stress-energy tensor. 

Finally, we remark that although complicated, the derived
stress-energy tensor may be employed as a source term of the
semi-classical Einstein field equations. Indeed, preliminary
calculations indicate that it is possible to find an
analytical perturbative solution to the back reaction, and
what makes this issue even more interesting and worth
further studies is the regularity of the geometry of the ABG
black hole. We intend to return to this group of problems
elsewhere.

\clearpage 

\def\cprime{$'$}

%%%%%%%%% Figure 1 %%%%%%%%%%%%%%
\begin{figure}[tbp]
\centering
\includegraphics{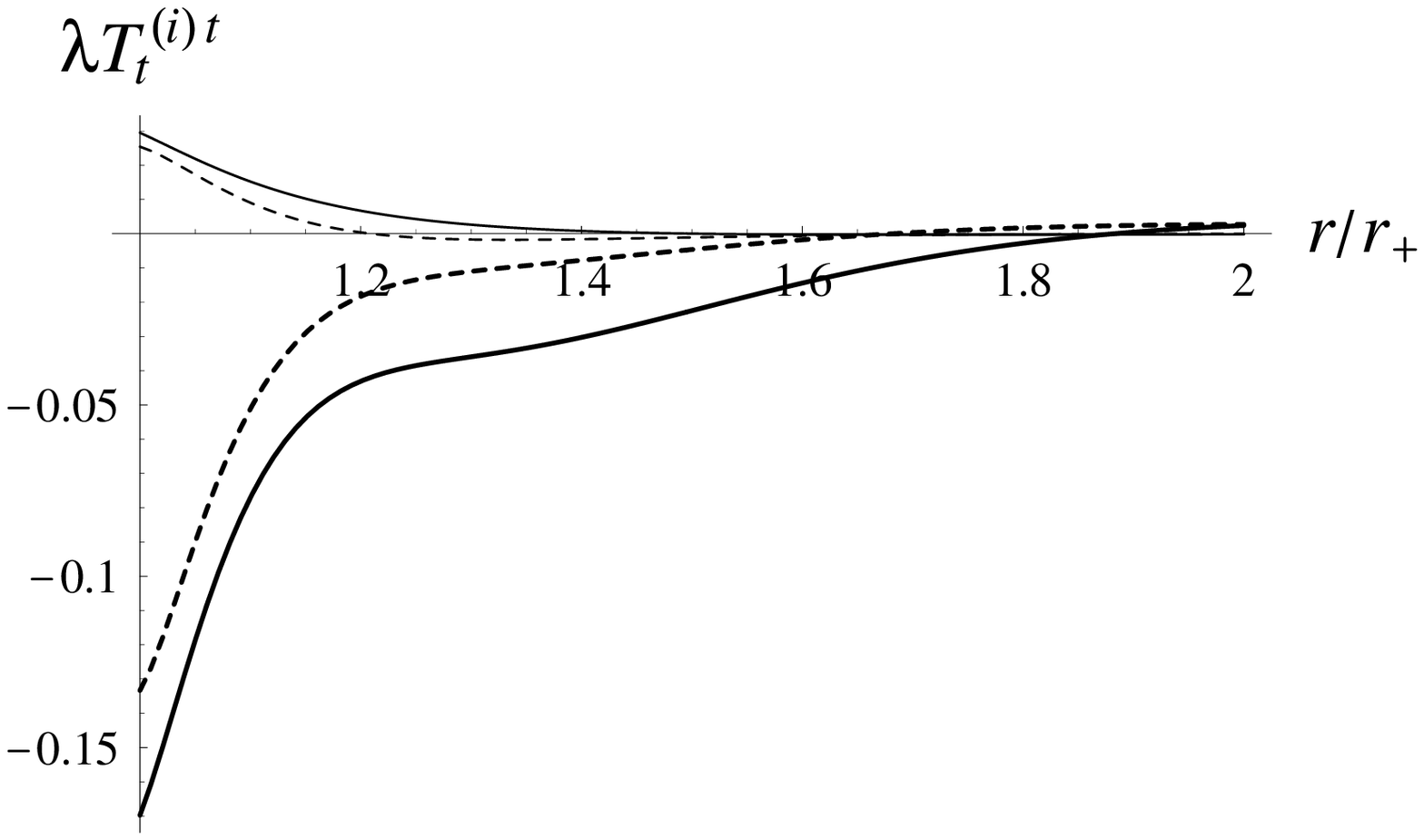}
\caption{The radial dependence of the rescaled time component
$\lambda T^{(1)t}_{t}$ ($\protect\lambda = 96\protect\pi^{2} M^{6}m^{2}$)
(solid thick line) and $\lambda T^{(0)t}_{t}$ (solid thin line)
contributing to the renormalized stress-energy tensor 
of the massive scalar field for the extremal ABG geometry.
The dashed lines correspond to the appropriate tensors for extremal 
RN black hole: $\lambda C^{t}_{t}$ (thin curve) and $\lambda D^{t}_{t}$
(thick curve). }
\end{figure}

%%%%%%%%% Figure 2 %%%%%%%%%%%%%%
\begin{figure}[tbp]
\centering
\includegraphics{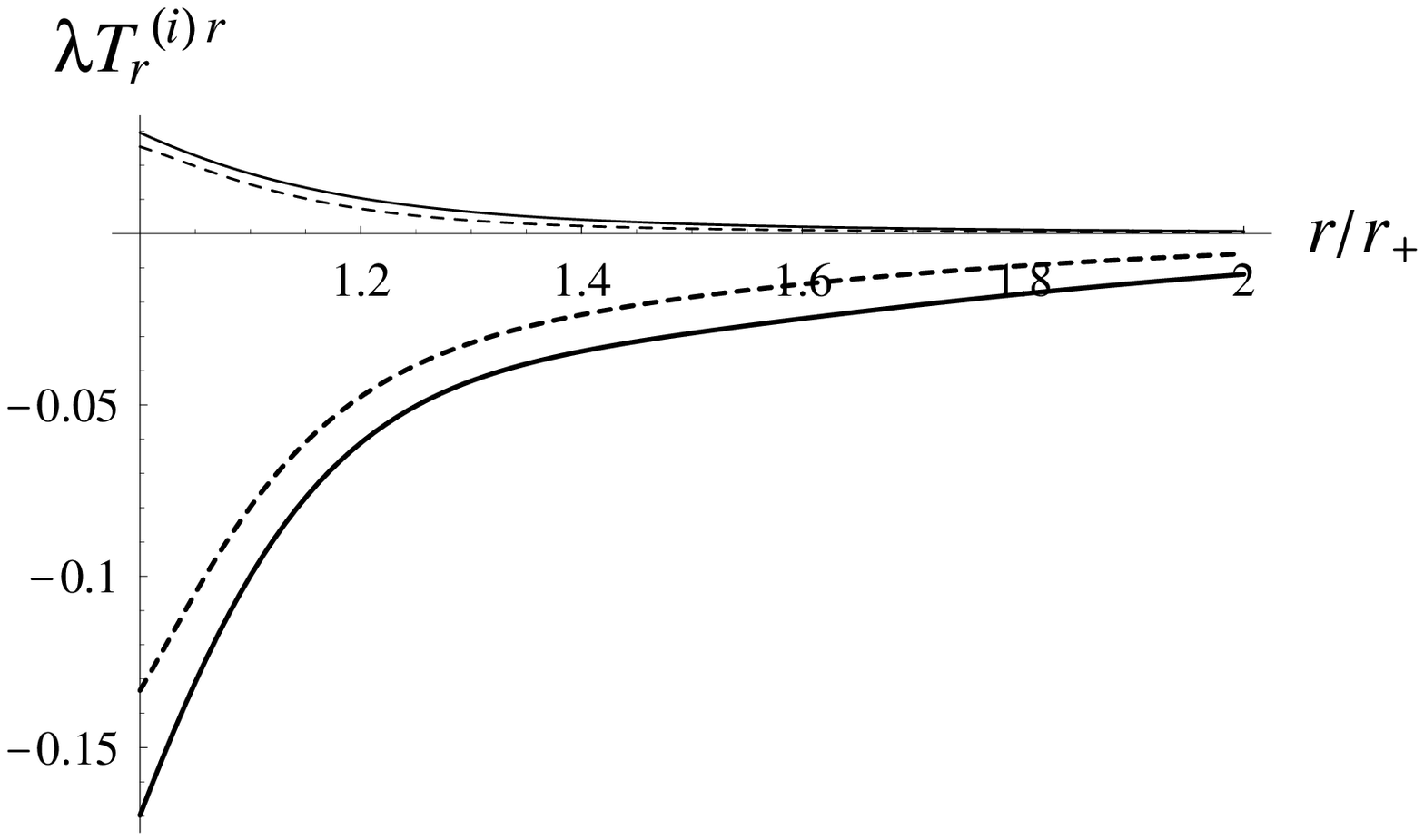}
\caption{The radial dependence of the rescaled radial component
$\lambda T^{(1)r}_{r}$ ($\protect\lambda = 96\protect\pi^{2} M^{6}m^{2}$)
(solid thick line) and $\lambda T^{(0)r}_{r}$ (solid thin line)
contributing to the renormalized stress-energy tensor 
of the massive scalar field for the extremal ABG geometry.
The dashed lines correspond to the appropriate tensors for extremal 
RN black hole: $\lambda C^{r}_{r}$ (thin curve) and $\lambda D^{r}_{r}$
(thick curve). }
\end{figure}

%%%%%%%%% Figure 3 %%%%%%%%%%%%%%
\begin{figure}[tbp]
\centering
\includegraphics{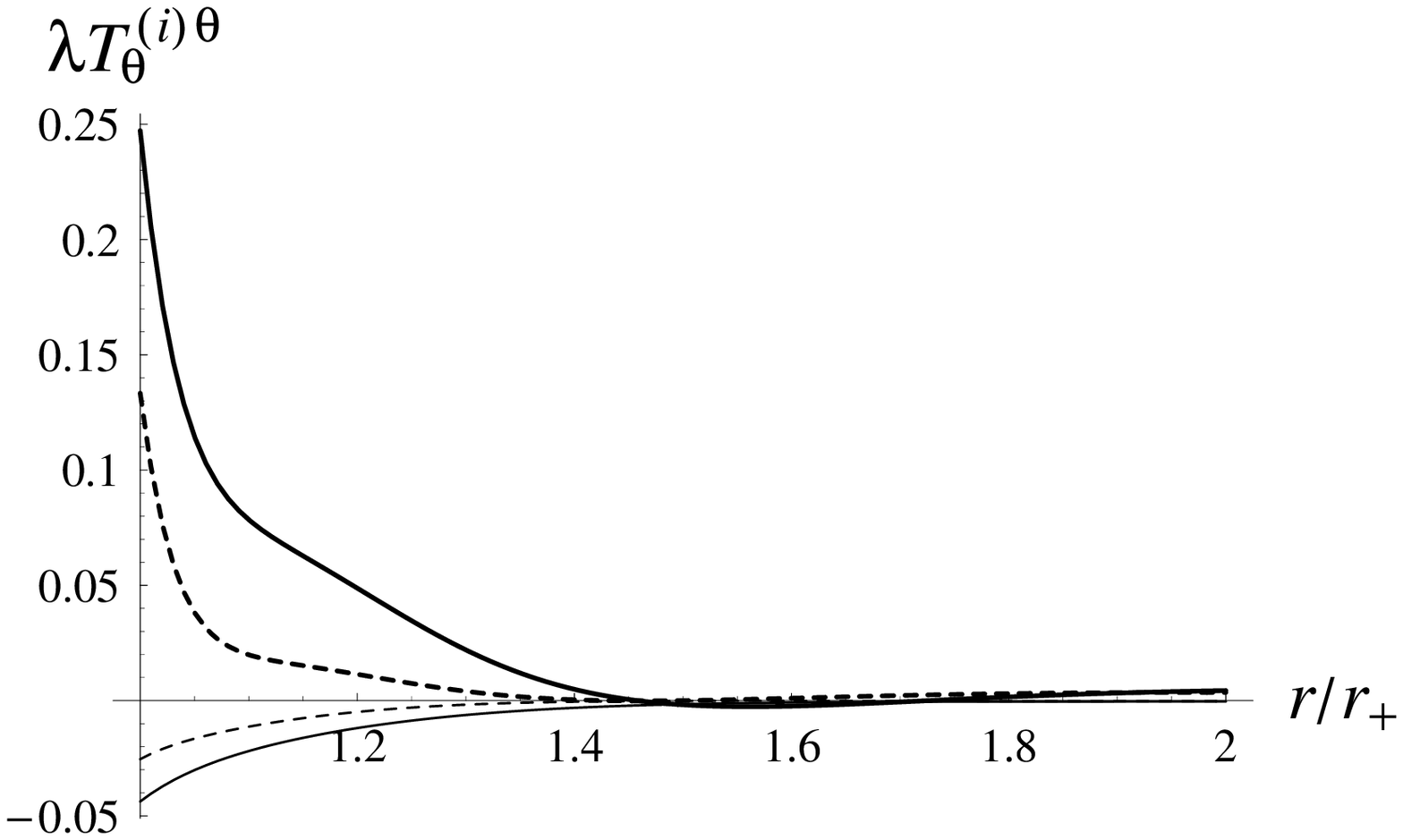}
\caption{The radial dependence of the rescaled angular component
$\lambda T^{(1)\theta}_{\theta}$ 
($\protect\lambda = 96\protect\pi^{2} M^{6}m^{2}$)
(solid thick line) and $\lambda T^{(0)\theta}_{\theta}$ (solid thin line)
contributing to the renormalized stress-energy tensor 
of the massive scalar field for the extremal ABG geometry.
The dashed lines correspond to the appropriate tensors for extremal 
RN black hole: $\lambda C^{\theta}_{\theta}$ (thin curve) 
and $\lambda D^{\theta}_{\theta}$
(thick curve). }
\end{figure}

%%%%%%%%% Figure 4 %%%%%%%%%%%%%%
\begin{figure}[tbp]
\centering
\includegraphics{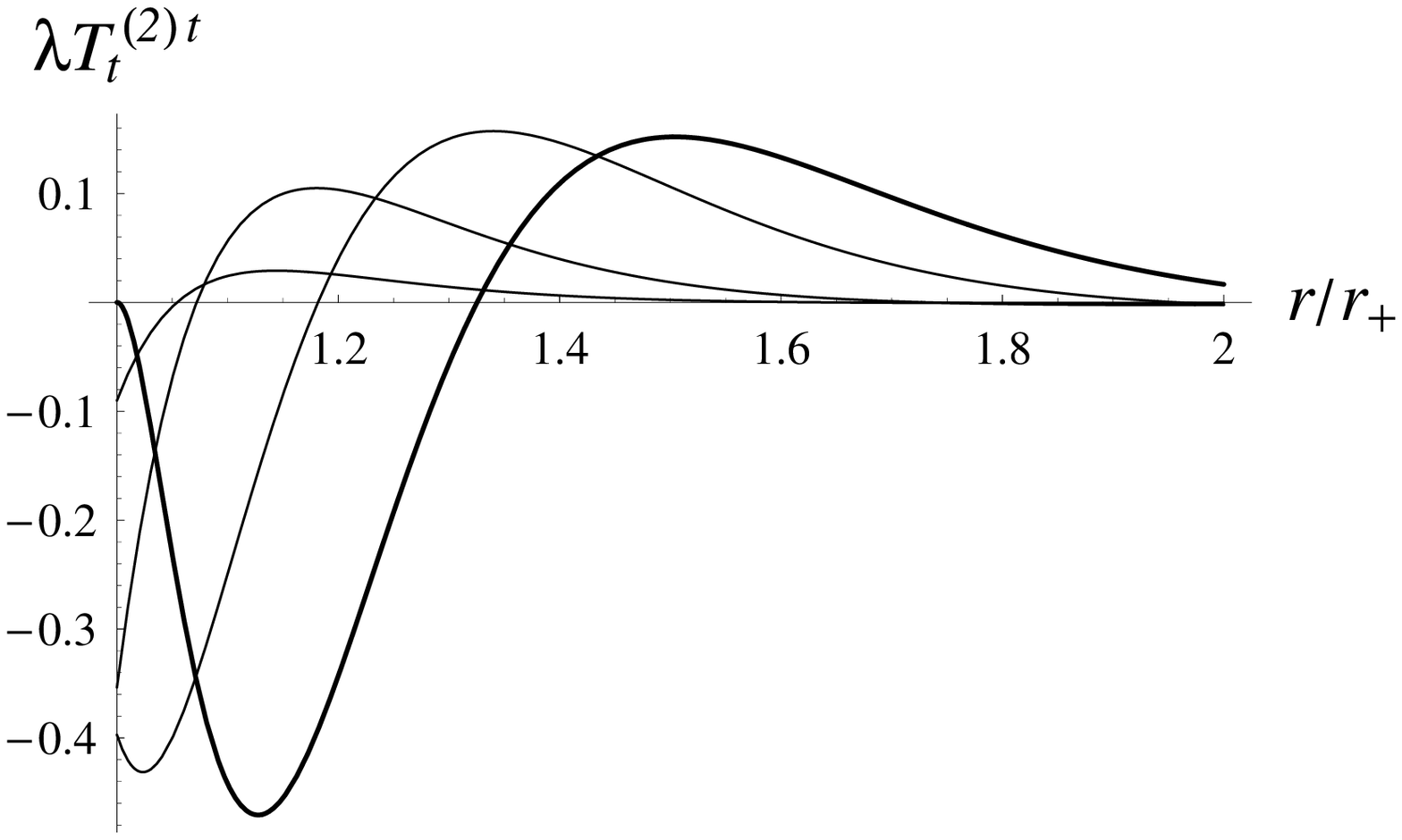}
\caption{The radial dependence of the rescaled time component 
$\lambda T^{(2)t}_{t}$ ($\protect\lambda = 96\protect\pi^{2} M^{6}m^{2}$)
contributing to the renormalized stress-energy tensor 
of the arbitrarily coupled massive scalar field in the ABG geometry.
The thick curve corresponds to the extremal case and the thin curves are for
a series of $q$ values approaching the extremality limit, $q=0.9, 1, 1.05$. }
\end{figure}

%%%%%%%%% Figure 5 %%%%%%%%%%%%%%
\begin{figure}[tbp]
\centering
\includegraphics{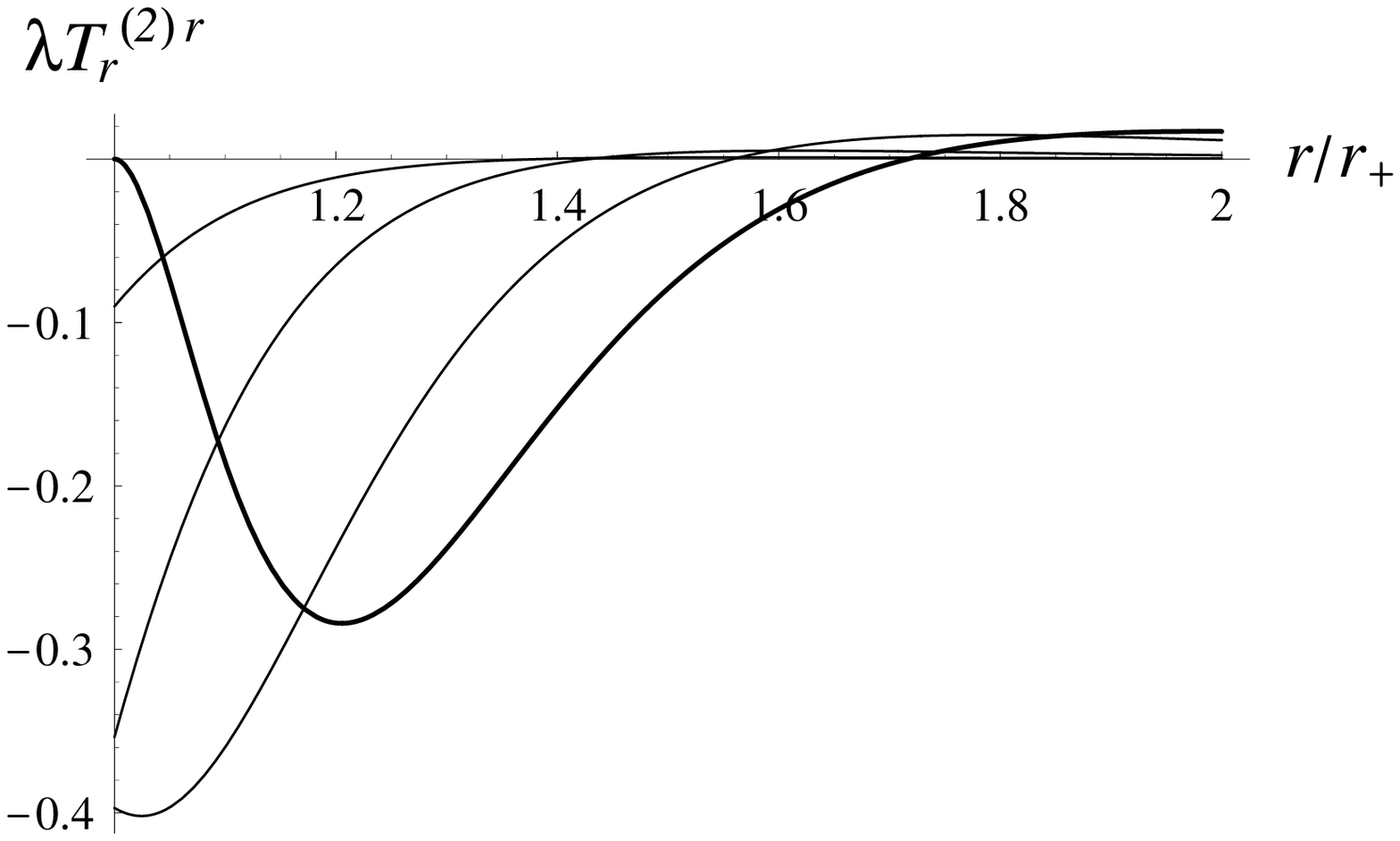}
\caption{ The radial dependence of the rescaled radial component 
$\lambda T^{(2)r}_{r}$ ($\protect\lambda = 96\protect\pi^{2} M^{6}m^{2}$)
contributing to the renormalized stress-energy tensor 
of the arbitrarily coupled massive scalar field in the ABG geometry.
The thick curve corresponds to the extremal case and the thin curves are for
a series of $q$ values approaching the extremality limit, $q=0.9, 1, 1.05$. }
\end{figure}

%%%%%%%%% Figure 6 %%%%%%%%%%%%%%
\begin{figure}[tbp]
\centering
\includegraphics{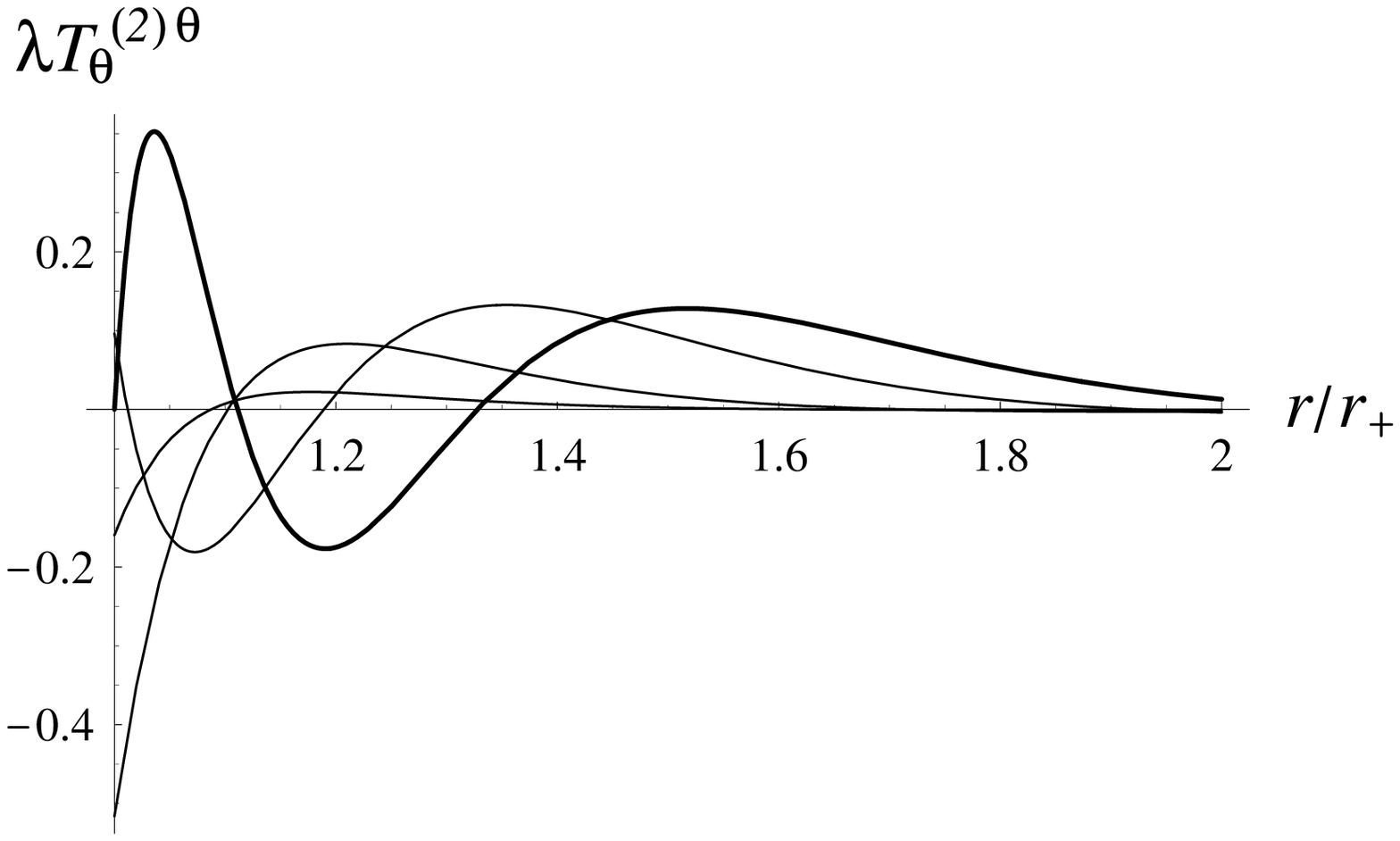}
\caption{The radial dependence of the rescaled angular component 
$\lambda T^{(2)\theta}_{\theta}$ 
($\protect\lambda = 96\protect\pi^{2} M^{6}m^{2}$)
contributing to the renormalized stress-energy tensor 
of the arbitrarily coupled massive scalar field in the ABG geometry.
The thick curve corresponds to the extremal case and the thin curves are for
a series of $q$ values approaching the extremality limit, $q=0.9, 1, 1.05$. }
\end{figure}

%%%%%%%%% Figure 7 %%%%%%%%%%%%%%
\begin{figure}[tbp]
\centering
\includegraphics{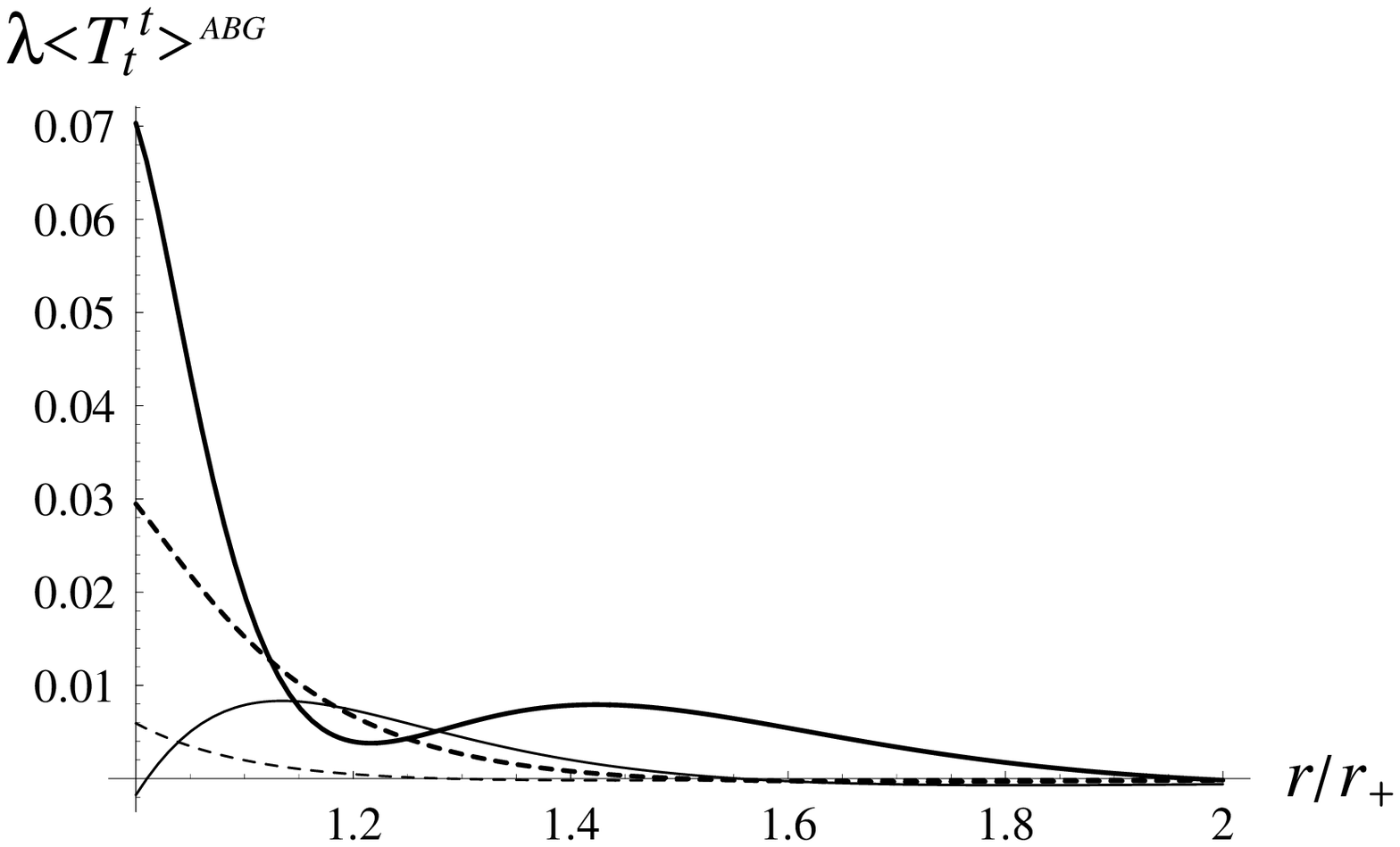}
\caption{The radial dependence of the rescaled time component 
$\lambda\langle T^{t}_{t}\rangle^{ABG}$ 
($\protect\lambda = 96\protect\pi^{2} M^{6}m^{2}$)
of the renormalized stress-energy tensor of 
the minimally coupled massive scalar field in the ABG geometry
solid lines) as compared to the
case of conformal coupling in this geometry (dashed lines). The thin curves
are for $q=1.02$ and the thick curves are for the extremal case. }
\end{figure}

%%%%%%%%% Figure 8 %%%%%%%%%%%%%%
\begin{figure}[tbp]
\centering
\includegraphics{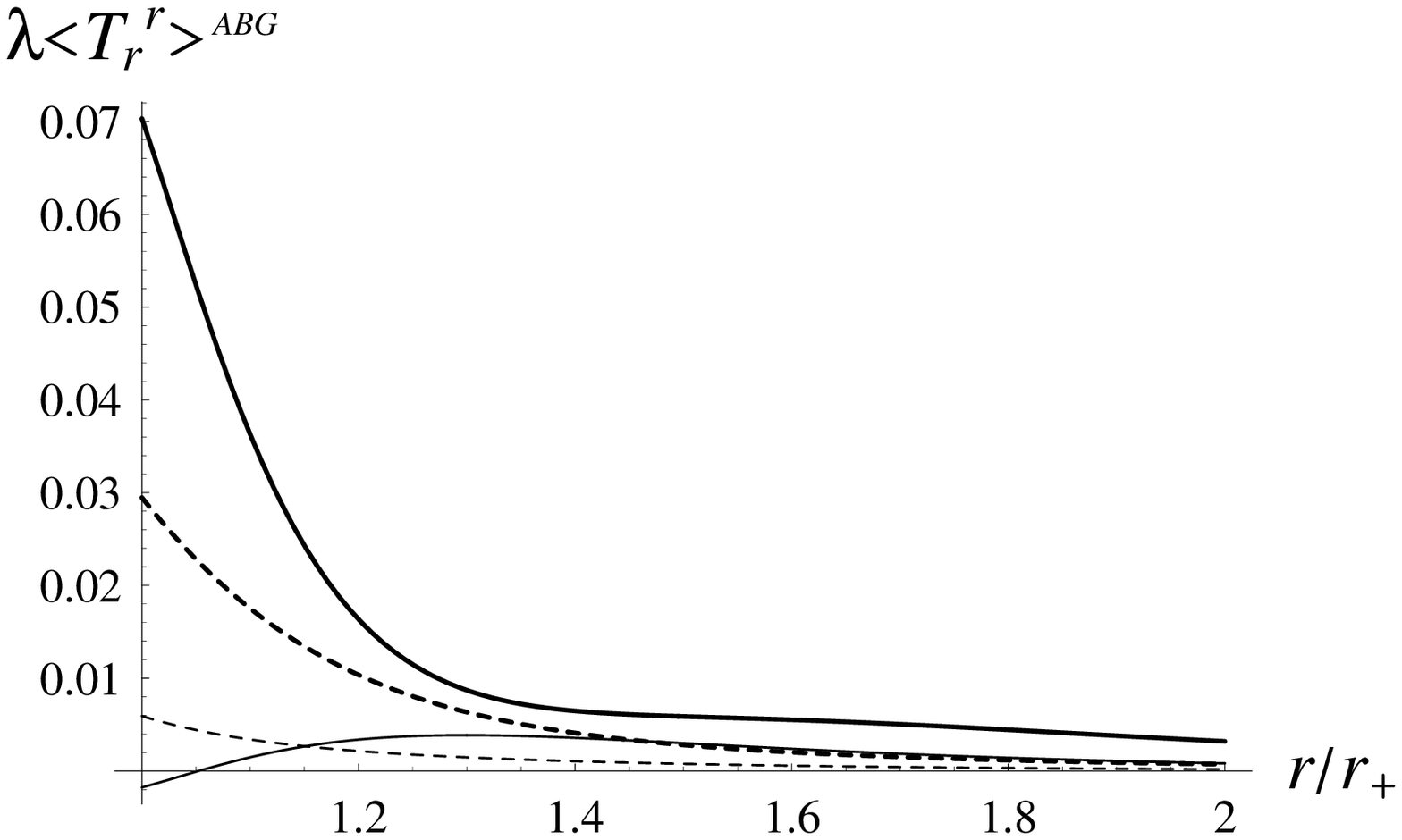}
\caption{The radial dependence of the rescaled radial component 
$\lambda\langle T^{r}_{r}\rangle^{ABG}$ 
($\protect\lambda = 96\protect\pi^{2} M^{6}m^{2}$)
of the renormalized stress-energy tensor of the 
minimally coupled massive scalar field in the ABG geometry
(solid lines) as compared to the
case of conformal coupling in this geometry (dashed lines). The thin curves
are for $q=1.02$ and the thick curves are for the extremal case. }
\end{figure}

%%%%%%%%% Figure 9 %%%%%%%%%%%%%%
\begin{figure}[tbp]
\centering
\includegraphics{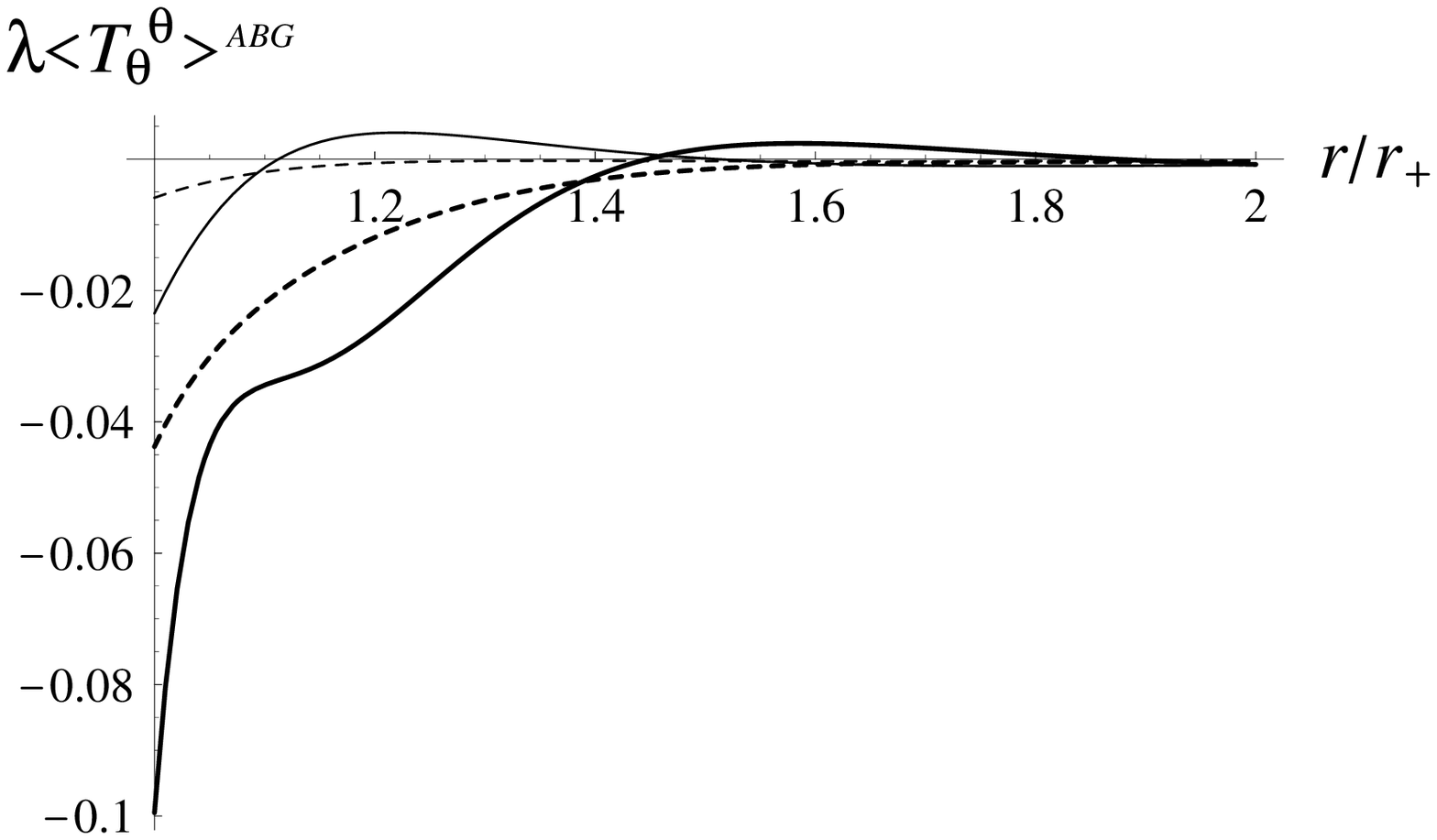}
\caption{The radial dependence of the rescaled angular component 
$\lambda\langle T^{\protect\theta}_{\protect\theta}\rangle^{ABG}$ 
($\protect\lambda = 96\protect\pi^{2} M^{6}m^{2}$)
of the renormalized stress-energy tensor of the minimally coupled massive 
scalar field in the ABG geometry (solid lines) 
as compared to the case of conformal coupling in this geometry
(dashed lines). The thin curves are for $q=1.02$ and the thick curves are
for the extremal case. }
\end{figure}

\end{document}